\documentclass{emulateapj}
\usepackage{apjfonts}

\usepackage{amsmath}

\usepackage{graphics}

\newcommand{\bd}{\begin{displaymath}}
\newcommand{\ed}{\end{displaymath}}

\slugcomment{accepted by ApJ}

\shorttitle{Evolution of massive black holes}

\begin{document}

\title{Cosmological evolution of massive black holes: effects of Eddington ratio distribution and quasar lifetime}

\author{Xinwu Cao\altaffilmark{1}}

\begin{abstract}

A power-law time-dependent lightcurve for active galactic nuclei
(AGNs) is expected by the self-regulated black hole growth scenario,
in which the feedback of AGNs expels gas and shut down accretion.
This is also supported by the observed power-law Eddington ratio
distribution of AGNs. At high redshifts, the AGN life timescale is
comparable with (or even shorter than) the age of the universe,
which set a constraint on the minimal Eddington ratio for AGNs on
the assumption of a power-law AGN lightcurve. The black hole mass
function (BHMF) of AGN relics is calculated by integrating the
continuity equation of massive black hole number density on the
assumption of the growth of massive black holes being dominated by
mass accretion with a power-law Eddington ratio distribution for
AGNs. The derived BHMF of AGN relics at $z=0$ can fit the measured
local mass function of the massive black holes in galaxies quite
well, provided the radiative efficiency $\sim 0.1$ and a suitable
power-law index for the Eddington ratio distribution are adopted. In
our calculations of the black hole evolution, the duty cycle of AGN
should be less than unity, which requires the quasar life timescale
$\tau_{\rm Q}\ga 0.5$ giga-years.
\end{abstract}


\altaffiltext{1}{Key Laboratory for Research in Galaxies and
Cosmology, Shanghai Astronomical Observatory, Chinese Academy of
Sciences, 80 Nandan Road, Shanghai, 200030, China; cxw@shao.ac.cn}

\section{Introduction}

There is evidence that most nearby galaxies contain massive black
holes at their centers, and the central massive black hole mass is
found to be tightly correlated with the velocity dispersion of the
galaxy \citep{2000ApJ...539L...9F,2000ApJ...539L..13G}, or the
luminosity of the spheroid component of its host galaxy
\citep*[e.g.,][]{1998AJ....115.2285M,2003ApJ...589L..21M}. These
correlations of the black hole mass with velocity dispersion/host
galaxy luminosity were widely used to estimate black hole masses,
and to derive the mass functions of the central massive black holes
in galaxies
\citep*[e.g.,][]{2002MNRAS.335..965Y,2004MNRAS.351..169M,2004MNRAS.354.1020S,2006MNRAS.365..134T,2007MNRAS.378..198G,2009NewAR..53...57S}.
On the other hand, quasars are powered by accretion onto massive
black holes, and the growth of massive black holes could be
dominantly governed by mass accretion in quasars. The massive black
holes are therefore the active galactic nuclei (AGN) relics
\citep{1982MNRAS.200..115S}, and the luminosity functions (LF) of
AGNs provide important clues on the growth of massive black holes.
The black hole mass function (BHMF) of AGN relics can be calculated
by integrating the continuity equation of massive black hole number
density on the assumption of the growth of massive black holes being
dominated by mass accretion, in which the activity of massive black
holes is described by the AGN LF
\citep*[e.g.,][]{1971ApJ...170..223C,1982MNRAS.200..115S,
1992MNRAS.259..421C,1992MNRAS.259..725S,2009ApJ...704...89S,2009ApJ...699..513L}. There
are two free parameters: the radiative efficiency $\eta_{\rm rad}$
and the mean Eddington ratio $\lambda=L_{\rm bol}/L_{\rm Edd}$ for
AGNs, adopted in most of such calculations on the cosmological
evolution of massive black holes
\citep*[e.g.,][]{2002MNRAS.335..965Y,2004MNRAS.351..169M,2004MNRAS.354.1020S}.
The derived BHMF of AGN relics in this way is required to match the
measured local BHMF at redshift $z=0$ by tuning the values of two
parameters $\eta_{\rm rad}$ and $\lambda$, which usually requires
almost all AGNs to be accreting close to the Eddington limit
\citep*[e.g.,][]{2002MNRAS.335..965Y,2004MNRAS.351..169M,2004MNRAS.354.1020S}.

In principle, the mean Eddington ratio for AGNs $\lambda$ is not a
free parameter, which can be estimated from a sample of AGNs with
measured black hole masses. One of the most effective approaches for
measuring masses of black holes in AGNs is the reverberation mapping
method \citep{1993PASP..105..247P,2000ApJ...533..631K}. Using the
tight correlation between the size of the broad-line region and the
optical luminosity established with the reverberation mapping method
for a sample of AGNs, the black hole masses of AGNs can be easily
estimated from their optical luminosity and width of broad emission
line. The mean Eddington ratio $\lambda\simeq 0.1$ at $z\sim 0.2$ to
$\simeq0.4$ at $z\sim 2$ was derived from a large sample of AGNs
with the analyses of the Sloan Digital Sky Survey (SDSS) by
\citet{2004MNRAS.352.1390M} \citep*[also
see][]{2004ApJ...608..136W,2010arXiv1006.3561K}.
\citet{2006ApJ...648..128K} pointed that the samples selected from
SDSS are heavily weighted toward high-luminosity objects due to the
limited sensitivity of SDSS. \citet{2006ApJ...648..128K} estimated
the Eddington ratios of AGNs discovered in the AGN and Galaxy
Evolution Survey (AGES), which is more sensitive than the SDSS. The
derived Eddington ratio distribution at fixed luminosity is well
described by a single log-normal distribution peaked at $\sim 0.25$
\citep*[also see][]{2008ApJ...680..169S,2009ApJ...700...49T}.
However, some other investigations showed that the Eddington ratios
of local AGNs spread over several orders of magnitude
\citep*[e.g.,][]{2002ApJ...564..120H,2006ApJ...643..641H}. The
Eddington ratio distribution for AGNs exhibits a power-law
distribution with an exponential cutoff at a high Eddington ratio
\citep{2008MNRAS.388.1011M,2009ApJ...698.1550H}, or a power-law distribution
with an additional log-normal component \citep{2009MNRAS.397..135K}.
Such a power-law Eddington ratio distribution is qualitatively
consistent with the self-regulated black hole growth scenario, in
which the feedback of AGN expels gas and shut down accretion
\citep*[e.g.,][]{2005ApJ...630..716H,2005ApJ...625L..71H,2009ApJ...698.1550H}.
This means that an AGN with bolometric luminosity $L_{\rm bol}$ may
contain a relatively small black hole accreting at a high rate or a
more massive black hole accreting at a lower rate. In this work, we
adopt a power-law Eddington ratio distribution with an exponential
cutoff at a high ratio to derive the BHMF of AGN with a LF, with
which the continuity equation for black hole number density is
integrated to calculate the cosmological evolution of BHMF of AGN
relics. The resultant BHMF of AGN relics is constrained by the
measured local BHMF. The conventional cosmological parameters
$\Omega_{\rm M}=0.3$, $\Omega_{\Lambda}=0.7$, and $H_0=70~ {\rm
km~s^{-1}~Mpc^{-1}}$ have been adopted in this work.

\section{The Eddington ratio distribution for active galactic
nuclei}

\citet{2009ApJ...698.1550H} suggested that the quasar lightcurve can
be described by
\begin{equation}
{\frac {dt}{d\log\lambda}}=\tau_{\rm
Q}\left({\frac{\lambda}{\lambda_{\rm peak}}}
\right)^{-\beta_l}\exp\left(-{\frac{\lambda}{\lambda_{\rm
peak}}}\right), \label{lightcurve}
\end{equation}
where $\lambda=L_{\rm bol}/L_{\rm Edd}$ ($L_{\rm Edd}=1.3\times
10^{38}M_{\rm bh}/M_\odot~{\rm ergs}~{\rm s}^{-1}$), $\tau_{\rm Q}$
is the quasar life timescale, and the parameter $\lambda_{\rm peak}$
describes the peak luminosity of quasars. This power-law lightcurve
is consistent with the self-regulated black hole growth model, in
which feedback produces a self-regulating ¡°decay¡± or ¡°blowout¡±
phase after the AGN reaches some peak luminosity and begins to expel
gas and shut down accretion \citep*[e.g.,][]{2005ApJ...630..716H}.
This lightcurve can be translated to an observed Eddington ratio
distribution $\zeta(\lambda)$,
\begin{equation}
\zeta(\lambda)={\frac
{d\mathcal{N}}{\mathcal{N}d\log\lambda}}=C_{l}\left({\frac{\lambda}{\lambda_{\rm
peak}}} \right)^{-\beta_l}\exp\left(-{\frac{\lambda}{\lambda_{\rm
peak}}}\right), \label{eddrat_dis}
\end{equation}
where $C_{l}$ is the normalization, if the switch-on of AGN activity
is balanced with switch-off of AGN activity and $\tau_{\rm Q}$ is
significantly shorter than the age of the universe at redshift $z$.
The power-law Eddington ratio distribution is consistent with those
derived with samples of nearby AGNs
\citep*[e.g.,][]{2004ApJ...613..109H,2005ApJ...634..901Y,2009ApJ...698.1550H}.
Such an Eddington ratio distribution has a lower cutoff at
$\lambda=\lambda_{\rm min,0}$, below which the sources are no longer
regarded as AGNs. In this work, we adopt $\lambda=\lambda_{\rm
min,0}=10^{-4}$ in all the calculations.

At high redshifts, the quasar life timescale $\tau_{\rm Q}$ is
comparable with (or even shorter than) the age of the universe at
redshift $z$. In this case, the time after the birth of the first
quasars is so short that most of them are still very luminous (i.e.,
accreting at high rates), and the lower limit on the Eddington
ratios for AGNs should be significantly higher than $\lambda_{\rm
min,0}$. The first quasars are predicted to have formed at  $z_{\rm
fq}\sim 10$
\citep*[e.g.,][]{2001ApJ...552..459H,2003ApJ...596...34B}, with
which we can estimate the minimal Eddington ratio $\lambda_{\rm
min}^\prime$ for AGNs at redshift $z$ as
\begin{equation}
\int\limits_{\lambda_{\rm min}^\prime(z)} \tau_{\rm
Q}\left({\frac{\lambda}{\lambda_{\rm peak}}}
\right)^{-\beta_l}\exp\left(-{\frac{\lambda}{\lambda_{\rm
peak}}}\right) {d\log\lambda}=t(z), \label{lambda_min}
\end{equation}
where $t(z)$ is the age of the universe at $z$ measured from $z_{\rm
fq}=10$ when the first quasars formed. For simplicity, we adopt
$\lambda_{\rm min}=\max[\lambda_{\rm min,0},\lambda_{\rm
min}^\prime(z)]$ in all our calculations on the black hole
evolution. In Fig. \ref{fig_eddrat}, we plot the minimal Eddington
ratios as functions of redshift $z$ for different quasar life
timescales $\tau_{\rm Q}$.

\vskip 1cm

\figurenum{1}
\centerline{\includegraphics[angle=0,width=8.5cm]{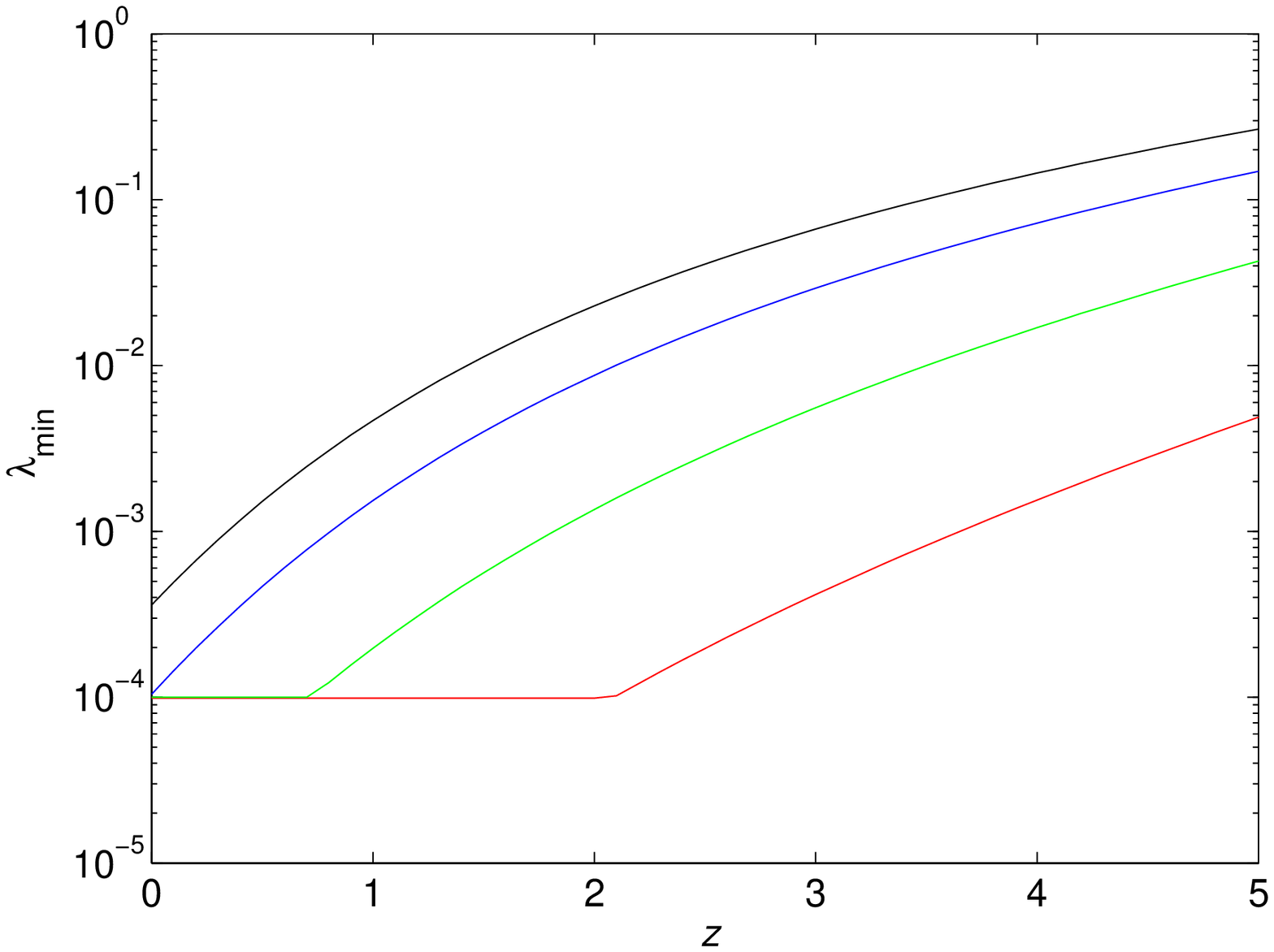}}
\figcaption{The minimal Eddington ratios as functions of redshift
$z$ for different AGN life timescales: $\tau_{\rm Q}=10^8$~(red),
$2.5\times 10^8$~(green), $5\times10^8$~(blue), and $7.5\times 10^8$
years (black), where $\lambda_{\rm min,0}=10^{-4}$ is adopted.
\label{fig_eddrat}   }\centerline{}

For a given black hole mass function $N_{\rm AGN}(M_{\rm bh},z)$ and
Eddington ratio distribution, the AGN LF $\Phi(z,L_{\rm bol})$ can
be calculated with
\begin{equation}
\Phi(z,L_{\rm bol})=\int\limits_{\lambda}N_{\rm AGN}(M_{\rm
bh},z){\frac {d\log M_{\rm bh}}{d\log L_{\rm bol}}}\zeta(\lambda)
d\log\lambda~~~{\rm Mpc^{-3}}(\log L_{\rm bol})^{-1}, \label{agn_lf}
\end{equation}
where $M_{\rm bh}/M_\odot=L_{\rm bol}/(1.3\times10^{38}\lambda)$,
$N_{\rm AGN}(M_{\rm bh},z)$ is the AGN BHMF at $z$, and the
Eddington ratio distribution $\zeta(\lambda)$ is given by Eq.
(\ref{eddrat_dis}). In this work, we assume that the AGN black hole
mass function has the same form as the AGN LF,
\begin{equation}
N_{\rm AGN}(M_{\rm bh},z)={\frac {N_{\rm AGN,0}(z)}{(M_{\rm
bh}/M_{\rm bh}^*)^{\beta_{\rm m1}}+(M_{\rm bh}/M_{\rm
bh}^*)^{\beta_{\rm m2}}}}~~~{\rm Mpc^{-3}}(\log M_{\rm bh})^{-1},
\label{agn_bhmf}
\end{equation}
where the parameters $N_{\rm AGN,0}$, $M_{\rm bh}^*$, $\beta_{\rm
m1}$ and $\beta_{\rm m2}$ are to be determined. Substituting Eq.
(\ref{agn_bhmf}) into Eq. (\ref{agn_lf}), we can calculate the AGN
LF with a given Eddington ratio distribution $\zeta(\lambda)$
provided the values of the four parameters in the AGN BHMF are
specified. In this work, we tune the values of these parameters till
the observed LF can be well reproduced by that calculated with Eq.
(\ref{agn_lf}). We adopt the LF given by
\citet{2007ApJ...654..731H}, which is calculated by using a large
set of observed quasar luminosity functions in various wavebands,
from the IR through optical, soft and hard X-rays \citep*[see][for
the details]{2007ApJ...654..731H},
\begin{equation}
\Phi(L_{\rm bol},z)=\frac {\phi_{*}}{(L_{\rm
bol}/L_{*})^{\gamma_1}+(L_{\rm bol}/L_{*})^{\gamma_2}}~~~{\rm
Mpc^{-3}}(\log L_{\rm bol})^{-1}, \label{lf}
\end{equation}
with normalization $\phi_{*}$,  break luminosity $L_{*}$, faint-end
slope $\gamma_1$, and bright-end slope $\gamma_2$. The break
luminosity $L_{*}$ evolves with redshift is given by
\begin{equation}
\log L_{*}=(\log
L_{*})_{0}+k_{L,1}\xi+k_{L,2}\xi^{2}+k_{L,3}\xi^{3},
\end{equation}
and the two slopes $\gamma_1$  and $\gamma_2$ evolves with redshift
as
\begin{equation}
\gamma_1=(\gamma_1)_{0}(\frac{1+z}{1+z_{\rm ref}})^{k_{\gamma_1}},
\end{equation}
and
\begin{equation}
\gamma_2=\frac{2(\gamma_2)_{0}}{(\frac{1+z}{1+z_{\rm
ref}})^{k_{{\gamma_2},1}}+(\frac{1+z}{1+z_{\rm
ref}})^{k_{{\gamma_2},2}}}.
\end{equation}
The parameter $\xi$ is
\begin{equation}
\xi=\log\left(\frac{1+z}{1+z_{\rm ref}}\right),
\end{equation}
and $z_{\rm ref}=2$ is fixed.  The other parameters of this LF are
as follows: $\log \phi_{*}({\rm Mpc}^{-3})=-4.825\pm 0.060$, $[\log
L_{*}(3.9\times 10^{33} {\rm erg~s}^{-1})]_{0}=13.036\pm 0.043$,
$k_{L,1}=0.632\pm 0.077$, $k_{L,2}=-11.76\pm 0.38$,
$k_{L,3}=-14.25\pm 0.80$, $(\gamma_1)_{0}=0.417\pm 0.055$,
$k_{\gamma_1}=-0.623\pm 0.132$, $(\gamma_2)_{0}=2.174\pm 0.055$,
$k_{{\gamma_2},1}=1.460\pm 0.096$, and $k_{{\gamma_2},2}=-0.793\pm
0.057$ \citep*[see][for the details]{2007ApJ...654..731H}. This LF
includes both the Compton-thin and thick sources.

We give some examples of the AGN BHMFs derived from an AGN LF with a
given Eddington ratio distribution at different redshits $z$ in Fig.
\ref {Fig_bhmf_lf}. It is found that the AGN LF can be well
reproduced by the calculations from the AGN BHMF with the form given
in Eq. (\ref{agn_bhmf}).

\vskip 1cm

\figurenum{2}
\centerline{\includegraphics[angle=0,width=8.5cm]{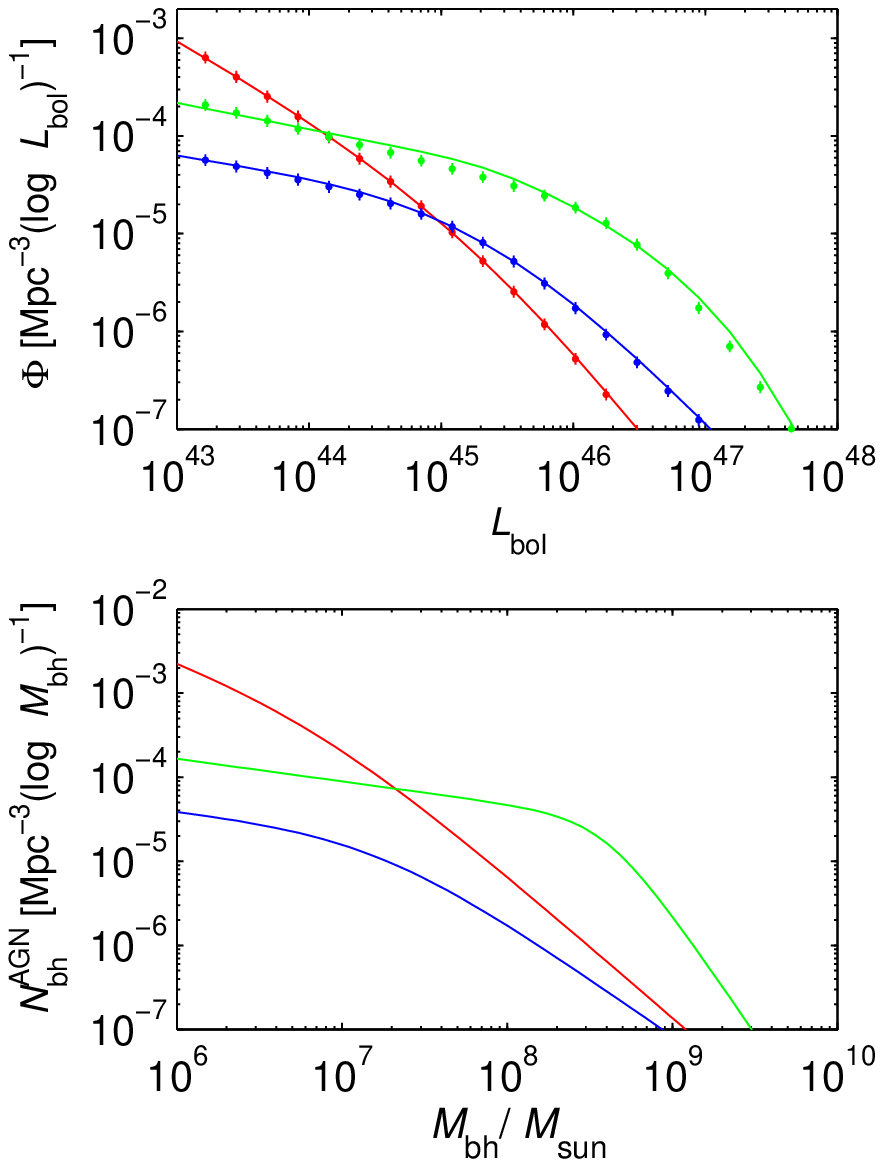}}
\figcaption{\textit{Top}: The comparison between the best-fitted LFs
(lines) calculated form the AGN BHMF with a given Eddington ratio
distribution and the LFs (dots) given by \citet{2007ApJ...654..731H}
at different redshifts: $z=0$~(red), $3$~(green), and $5$~(blue).
The parameters: $\beta_l=0.3$, $\lambda_{\rm peak}=2.5$, and
$\tau_{\rm Q}=5\times10^8$~years, are adopted. \textit{Bottom}: The
derived AGN BHMFs at different redshits: $z=0$~(red), $3$~(green),
and $5$~(blue). \label{Fig_bhmf_lf}  }\centerline{}

\section{The cosmological evolution of massive black holes}

{The evolution of massive black hole number density is described by
\citep{1992MNRAS.259..725S}
\begin{equation}
{\frac {\partial n(M_{\rm bh},t)}{\partial t}}+{\frac
{\partial}{\partial M_{\rm bh}}}[n(M_{\rm bh},t)<\dot{M}(M_{\rm
bh},t)]=0, \label{bhmf_evol_sb}
\end{equation}
where the black hole mass function $n(M_{\rm bh},t)$ is in units of
${\rm Mpc}^{-3}M_{\rm bh}^{-1}$, $<\dot{M}(M_{\rm bh},t)>$ is the
mean mass accretion rate for the black holes with $M_{\rm bh}$, and
the efforts of mergers of black holes are neglected \citep*[see][for
the detailed discussion]{2009ApJ...690...20S,2008MNRAS.390..561C}.
The black hole mass function $N(M_{\rm bh},t)$ used in this work is
in units of ${\rm Mpc}^{-3}(\log M_{\rm bh})^{-1}$, Equation
(\ref{bhmf_evol_sb}) can be re-written as
\begin{equation}
{\frac {1}{M_{\rm bh}}}{\frac {\partial N(M_{\rm bh},t)}{\partial
t}}+{\frac {\partial}{\partial M_{\rm bh}}}\left[{\frac {N(M_{\rm
bh},t)<\dot{M}(M_{\rm bh},t)>}{M_{\rm bh}}}\right]=0,
\label{bhmf_evol}
\end{equation}
since $N(M_{\rm bh},t)\equiv \ln10 M_{\rm bh}n(M_{\rm bh},t)$.}

It was suggested that the standard thin accretion disk will transit
to an advection dominated accretion flow (ADAF) when the
dimensionless mass accretion rate $\dot{m}$ is lower than a critical
value $\dot{m}_{\rm crit}$ ($\dot{m}=\dot{M}/\dot{M}_{\rm Edd}$;
$\dot{M}_{\rm Edd}=1.3\times 10^{38}M_{\rm bh}/0.1M_\odot c^2$)
\citep{1995ApJ...452..710N}. The radiative efficiency for ADAFs is
significantly lower than that for standard thin disks, and it
decreases with decreasing mass accretion rate $\dot{m}$. It was
suggested that the radiative efficiency $\eta_{\rm rad}$ can be
described with
\begin{equation}
 \eta_{\rm rad}=\left\{ \begin{array}{ll}
        \eta_{\rm rad,0}, & \mbox{if $\dot{m}\ge \dot{m}_{\rm crit}$};\\
         \eta_{\rm rad,0}\left({\frac {\dot{m}}{\dot{m}_{\rm crit}
}}\right)^s, & \mbox{if $\dot{m}<\dot{m}_{\rm crit}$ },\end{array}
\right. \label{etarad}
        \end{equation}
where $\dot{m}_{\rm crit}=0.01$ is adopted in this work
\citep*[see][for the detailed discussion and the references
therein]{2002luml.conf..405N}, and the parameter $s=1$ is suggested
in \citet{1995ApJ...452..710N}. The calculations of the ADAFs
surrounding rotating black holes in the general relativistic frame
showed that the value of $s$ is in the range of $\sim 0.2-1.1$
depending on the value of black hole spin parameter $a$
\citep{2010ApJ...716.1423X}. In all our calculations, we adopt
$s=0.5$ \citep*[e.g.,][]{2008MNRAS.388.1011M,2010ApJ...715L..99D}.
The mean mass accretion rate for the black holes with $M_{\rm bh}$
at redshift $z$ can be calculated with
\begin{displaymath}
N(M_{\rm bh},z)<\dot{M}(M_{\rm
bh},z)>~~~~~~~~~~~~~~~~~~~~~~~~~~~~~~~~~~~~~~~~~~~~~~~~~~~~~~~~~~~~~~~
\end{displaymath}
\begin{equation}
=\int\limits_{\lambda} {\frac {\zeta(\lambda)N_{\rm AGN}(M_{\rm
bh},z)(M_{\rm bh}/M_\odot)\lambda L_{\rm Edd,\odot}(1-\eta_{\rm
rad})}{\eta_{\rm rad}c^2}}d\log\lambda, \label{mdot_aver}
\end{equation}
where $N_{\rm AGN}(M_{\rm bh},z)$ is the AGN BHMF at $z$, and the
radiative efficiency $\eta_{\rm rad}$ is given by Eq.
(\ref{etarad}).

The black hole evolution equation can be rewritten as
\begin{equation}
{\frac {1}{M_{\rm bh}}}{\frac {\partial N(M_{\rm bh},z)}{\partial
z}}=-{\frac {dt}{dz}}{\frac {\partial}{\partial M_{\rm
bh}}}\left[{\frac {N(M_{\rm bh},t)<\dot{M}(M_{\rm bh},t)>}{M_{\rm
bh}}}\right]. \label{bhmf_evolz}
\end{equation}
As described in \S 2, the AGN BHMF $N_{\rm AGN}(M_{\rm bh},z)$ can
be calculated with a given Eddington ratio distribution
(\ref{eddrat_dis}) and the AGN LF (\ref{lf}), and the mean mass
accretion rate can be calculated with Eq. (\ref{mdot_aver}).
Integrating Eq. (\ref{bhmf_evolz}) over $z$ from $z_{\rm max}$ with
derived AGN BHMF $N_{\rm AGN}(M_{\rm bh},z)$ and suitable initial
conditions at $z_{\rm max}$, the cosmological evolution of massive
black holes is available. The duty cycle $\delta$ of AGN is defined
as
\begin{equation}
\delta(M_{\rm bh},z)={\frac {N_{\rm AGN}(M_{\rm bh},z)}{N(M_{\rm
bh},z)}},\label{duty_cyc}
\end{equation}
which is required to be less than unity. In all our calculations, we
assume that the duty cycle $\delta=0.5$ at $z_{\rm max}=5$. There
are three free parameters, $\eta_{\rm rad,0}$, $\beta_l$, and
$\lambda_{\rm peak}$, in our calculations, which are tuned to let
the BHMF of the AGN relics at $z=0$ fit the measured local BHMF
given in \citet{2009ApJ...690...20S}. This local BHMF encompasses
the range of several estimates of BHMF with different methods
\citep*{2003ApJ...585L.117B,2004MNRAS.351..169M,2004MNRAS.354.1020S,
2005SSRv..116..523F,2007ApJ...663...53T,2007ApJ...669...45H}. We
find that the final results are insensitive to the initial
conditions at $z_{\rm max}$, because the fraction of local black
hole mass accreted at very high redshifts can be neglected.

The results for the evolution of the total BHMFs and AGN BHMFs with
redshift $z$ are plotted in Fig. \ref{fig_bhmf_b} for different
values of the model parameters. In Fig. \ref{fig_dutycyc}, the duty
cycles $\delta$ of AGNs are plotted as functions of black hole mass
$M_{\rm bh}$ at different redshifts $z$. In this work, the AGNs
accreting at $\ge \dot{m}_{\rm crit}$ are referred as bright AGNs,
in which radiative efficiently accretion disks are present. We plot
the radiative efficiency evolving with redshift in Fig.
\ref{fig_eta}.

\vskip 1cm

\figurenum{3}
\centerline{\includegraphics[angle=0,width=8.5cm]{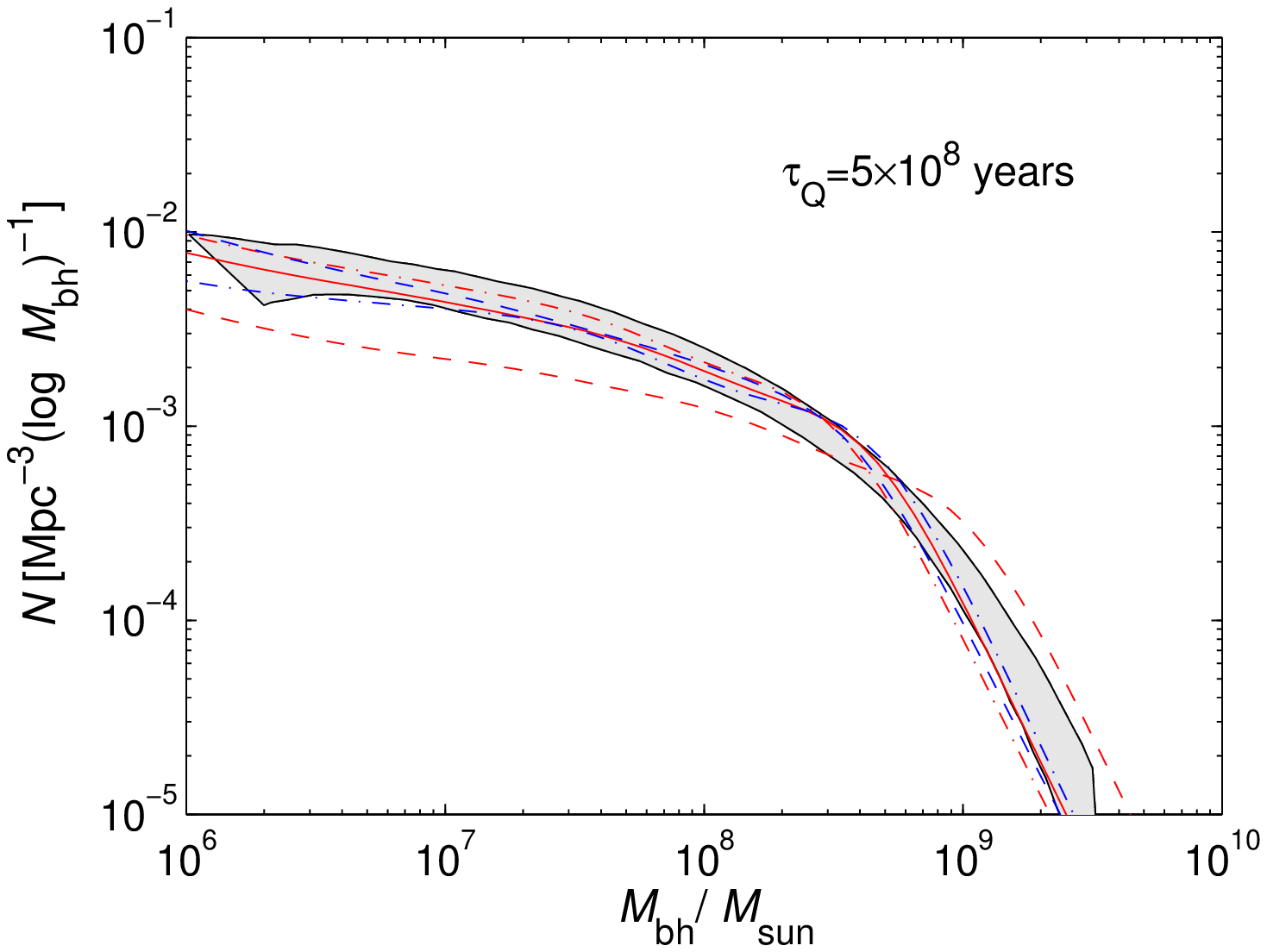}}
\figcaption{The local BHMFs at $z=0$. The shaded area encompasses
several estimates of the local BHMF \citep*[see Fig. 5 in][and
references therein]{2009ApJ...690...20S}. The lines represent the
BHMF of the AGN relics calculated with AGN life timescale $\tau_{\rm
Q}=5\times10^8$~years. The red solid lines are the results with
$\beta_l=0.3$ and $\lambda_{\rm peak}=2.5$, while the red dashed and
dash-dotted lines represent the results with $\lambda_{\rm peak}=1$
and $3.5$, respectively ($\beta_l=0.3$). The blue dashed and
dash-dotted lines represent the results with $\beta_l=0.2$ and
$0.4$, respectively ($\lambda_{\rm peak}=2.5$). \label{fig_bhmf_a}
}\centerline{}

\section{Discussion}

Our estimates of the lower limits on the Eddington ratios of AGNs
show that $\lambda_{\rm min}$ increases with redshift $z$, which
implies that the mean Eddington ratio is higher at high redshifts.
At high redshifts, the quasar life timescale $\tau_{\rm Q}$ is
comparable with (or even shorter than) the age of the universe at
redshift $z$, and most of the AGNs are therefore still very luminous
(i.e., accreting at high rates). The mean Eddington ratios for these
AGNs are relatively higher than those at low redshifts. The
estimates of the Eddington ratios for AGNs show that the mean
Eddington ratio increases with $z$
\citep*[e.g.,][]{2004MNRAS.352.1390M,2004ApJ...608..136W}, which is
qualitatively consistent with our results.

Unlike most of the previous works
\citep*[e.g.,][]{2002MNRAS.335..965Y,2004MNRAS.351..169M,2004MNRAS.354.1020S},
in which a single mean Eddington ratio is adopted as a free
parameter, we use an Eddington ratio distribution in our
calculations. Such a power-law Eddington ratio distribution for AGNs
is expected by the self-regulated black hole growth model
\citep{1998A&A...331L...1S,2005ApJ...630..716H}, which is also
supported by the Eddington ratio estimates for AGNs
\citep*[e.g.,][]{2004ApJ...613..109H,2005ApJ...634..901Y,2009ApJ...698.1550H,2009MNRAS.397..135K}.

\vskip 1cm

\figurenum{4}
\centerline{\includegraphics[angle=0,width=8.5cm]{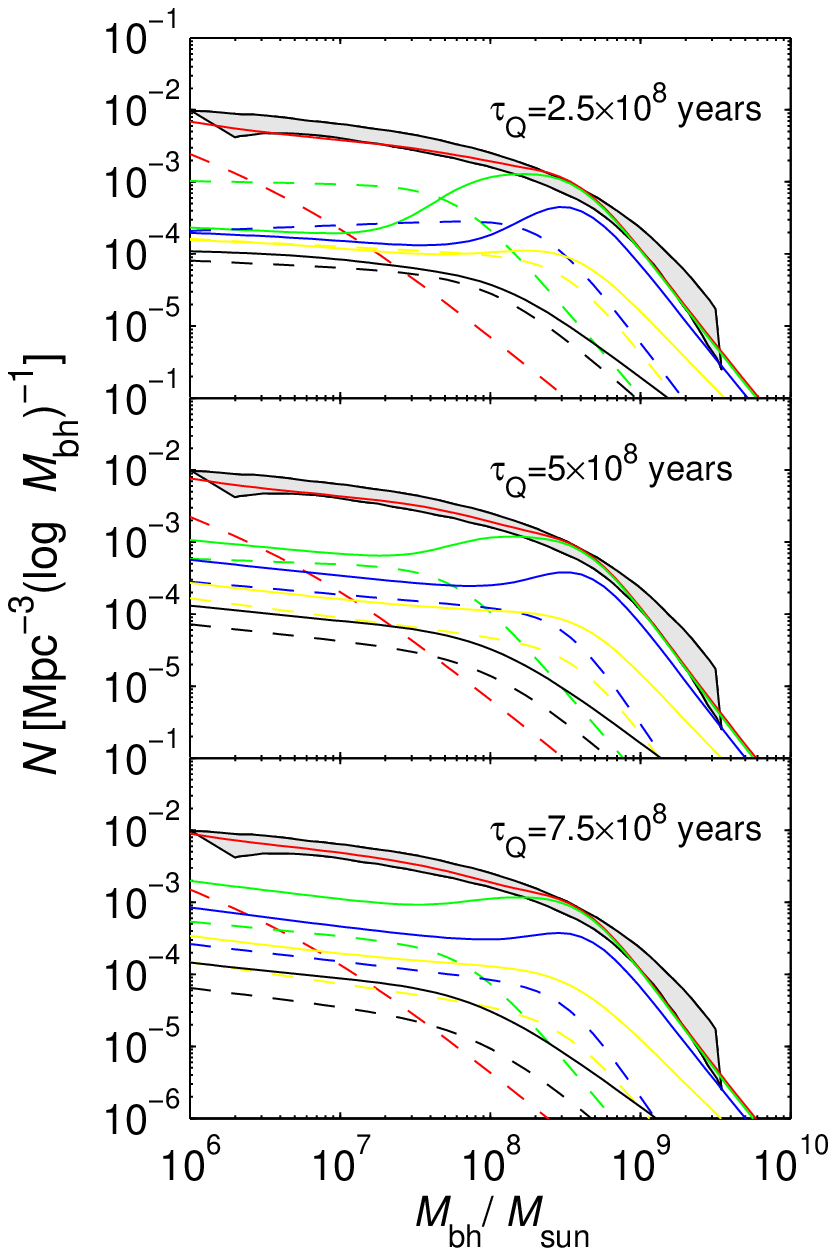}}
\figcaption{The BHMFs at different redshifts: $z=0$~(red),
$1$~(green), $2$~(blue), $3$~(yellow), and $4$~(black). The solid
lines represent the total BHMFs, while the dashed lines are for AGN
BHMFs. The parameters for the Eddington ratio distribution:
$\beta_l=0.3$ and $\lambda_{\rm peak}=2.5$, are adopted. The shaded
area is the measured local BHMF. In the upper panel, the results are
derived with AGN life timescale $\tau_{\rm Q}=2.5\times10^8$~years,
while the results with $\tau_{\rm Q}=5\times10^8$ and
$7.5\times10^8$~years are plotted in the middle and lower panels,
respectively. \label{fig_bhmf_b} }\centerline{}

There are three free parameters, $\eta_{\rm rad,0}$, $\beta_l$, and
$\lambda_{\rm peak}$, in our calculations on the evolution of
massive black holes.  The resultant local BHMF for the AGN relics is
sensitive to the value of the peak luminosity of AGNs. It is found
that the measured local BHMF can be well reproduced by the BHMF of
the AGN relics at $z=0$ calculated in this work, if the three
parameters: $\eta_{\rm rad,0}=0.11$, $\beta_l=0.3$, and
$\lambda_{\rm peak}=2.5$, are adopted (see Fig. \ref{fig_bhmf_a}).
\citet{2009ApJ...698.1550H} suggested that $\beta_l\simeq 0.3-0.8$
based on their self-regulated black hole growth model calculations,
and our calculations also provide a useful constraint on the value
of $\beta_l$. Our results show that the peak Eddington ratio of AGNs
$\sim 2.5$ is required for modeling the local BHMF, which implies
that a small fraction of AGNs are accreting at slightly
super-Eddington rates. This is consistent with the estimates for
different samples of AGNs
\citep*[e.g.,][]{2004ApJ...608..136W,2009MNRAS.398.1905W,2010ApJ...716L..31A,2010arXiv1006.1342W}.

\vskip 1cm

\figurenum{5}
\centerline{\includegraphics[angle=0,width=8.5cm]{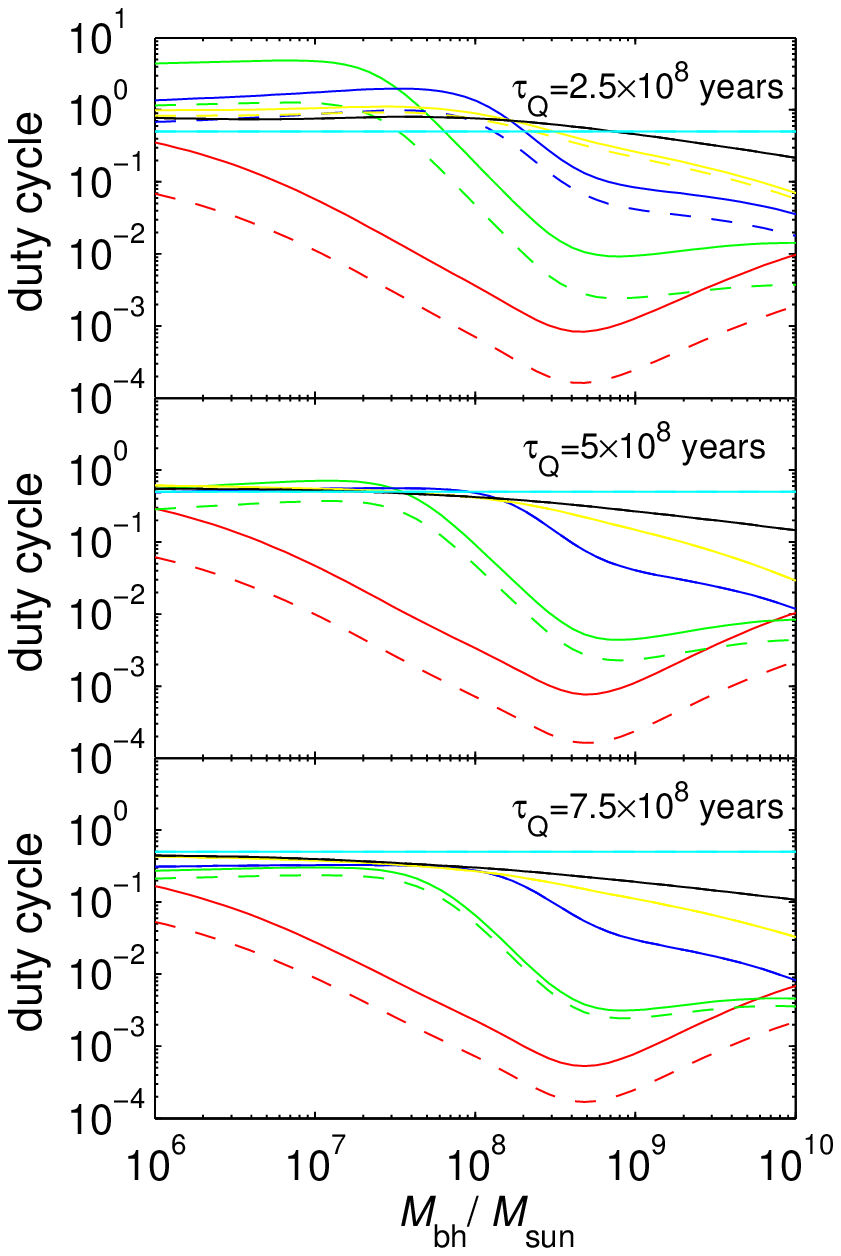}}
\figcaption{The AGN duty cycles as functions of black hole mass
$M_{\rm bh}$ at different redshifts: $z=0$(red), $1$(green),
$2$(blue), $3$(yellow), $4$(black), and $5$(cyan). The solid lines
represent the duty cycles of all AGNs (i.e., the sources with
$\lambda\ge\lambda_{\rm min}$), while the dashed lines represent the
duty cycles of bright AGNs (i.e., the sources accreting with
$\dot{m}\ge\dot{m}_{\rm crit}$). The parameters for the Eddington
ratio distribution: $\beta_l=0.3$, $\lambda_{\rm peak}=2.5$, are
adopted in the calculations. \label{fig_dutycyc}  }\centerline{}

The AGN BHMFs can be calculated from the AGN LF either with a
distribution of Eddington ratios or a single mean Eddington ratio.
The derived BHMF of AGN relics at redshift $z=0$ is required to
match the measured local BHMF, which usually leads to almost all
AGNs be accreting close to the Eddington limit in the calculations
with a single mean Eddington ratio. This is inconsistent with the
observations of AGN samples
\citep*[e.g.,][]{2002ApJ...564..120H,2004MNRAS.352.1390M,2004ApJ...608..136W,2006ApJ...643..641H,2006ApJ...648..128K,2010arXiv1006.3561K}.
The main difference between this work and the previous works is that
a power-law Eddington ratio distribution instead of a single mean
Eddington ratio is adopted in this work. The AGN BHMFs derived in
this work are larger than those calculated with a single Eddington
ratio close to unity adopted in those previous works
\citep*[e.g.,][]{2002MNRAS.335..965Y,2004MNRAS.351..169M,2004MNRAS.354.1020S}
especially at low redshifts, because the mean ratios for the
power-law Eddington ratio distributions used in this work are
significantly lower than unity. We find that the AGN BHMFs are
larger than the total BHMFs for relative small black holes at low
redshifts if the quasar life timescale $\tau_{\rm Q}\la 5\times
10^8$~years (see Fig. \ref{fig_bhmf_b}), i.e., the duty cycle for
AGNs $\delta>1$ (see Fig. \ref{fig_dutycyc}), which is unphysical.
These results imply that the quasar life timescale $\tau_{\rm Q}\ga
5\times 10^8$~years, which is qualitatively consistent with the
results derived either from theoretical calculations or
observational data
\citep*{2005Natur.433..604D,2009ApJ...698.1550H,2010arXiv1006.3561K}.
As an Eddington ratio distribution is adopted in our calculations,
the AGNs with black hole mass $M_{\rm bh}$ may be accreting at very
low rates. We also calculate the cosmological evolution of the duty
cycle $\delta_{\rm b}$ for the bright AGNs accreting at
$\dot{m}>\dot{m}_{\rm crit}$ as functions of $M_{\rm bh}$ (see Fig.
\ref{fig_dutycyc}). We find that the bright AGN duty cycles
$\delta_{\rm b}$ are significantly higher than $\delta_{\rm b}$ only
at low redshifts, while they converge at high redshifts ($z\ga 2$),
which is caused by the value of $\lambda_{\rm min}$ increasing with redshift $z$ (see Fig. \ref{fig_eddrat}).
It is found that the bright AGN duty cycles are
higher at high redshifts, and they decrease with increasing black
hole mass at low redshifts, which are qualitatively consistent with
those obtained in the previous works
\citep*[e.g.,][]{2004MNRAS.351..169M,2004MNRAS.353.1035M,2006ApJ...647L..17W,2009ApJ...690...20S}.
Our results show that the massive black holes were grown earlier
than their less massive counterparts (see Fig. \ref{fig_bhmf_b}),
which is similar to that found by \citet{2004MNRAS.353.1035M}. The
black holes with $M_{\rm bh}\la 10^8 M_\odot$ were dominantly grown
at redshifts $z\la 1$ (see Fig. \ref{fig_bhmf_b}).

{The Eddington ratios derived with bright AGN samples usually
exhibit a log-normal distribution
\citep{2006ApJ...648..128K,2008ApJ...680..169S,2010arXiv1006.3561K},
while those derived with the samples containing fainter sources show
a power-law distribution
\citep{2004ApJ...613..109H,2005ApJ...634..901Y,2009ApJ...698.1550H},
or a power-law distribution with an additional log-normal component
at high Eddington ratios \citep{2009MNRAS.397..135K}. The peak and
the dispersion of the log-normal Eddington ratio distribution were
found to be almost independent of the black hole mass and redshift
\citep{2006ApJ...648..128K}, or they have very weak dependence on
the black hole mass and redshift for a large AGN sample selected
from the Sloan Digital Sky Survey \citep{2008ApJ...680..169S}.
\citet{2008MNRAS.390..561C} calculated the evolution of massive
black holes using a log-normal Eddington ratio distribution derived
by \citet{2006ApJ...648..128K}, and found that the measured local
BHMF always cannot be reproduced by their model calculations unless
a black hole mass dependent radiative efficiency is assumed
\citep*[see][for the details]{2008MNRAS.390..561C}. The sources in a
large sample of nearby galaxies are separated into two populations
by the age of galaxies, which show a power-law and log-normal
Eddington ratio distributions respectively
\citep{2009MNRAS.397..135K}. The power-law Eddington ratio
distribution for the subsample exhibits a similar slope, $\sim
-0.8$, which is independent of black hole mass \citep*[see][for the
details]{2009MNRAS.397..135K}. In this work, a power-law Eddington
ratio distribution independent of black hole mass and redshift is
assumed in our calculations, which seems to be a reasonable
assumption. We note that the slope of the power-law Eddington ratio
distribution derived in this work is flatter than that of the
distribution derived with a subsample of nearby galaxies in
\citet{2009MNRAS.397..135K}. The observed Eddington ratio
distribution of all sources in their sample is a mixture of a
power-law and log-normal distributions, the slope of which in the
range of $\lambda \sim 10^{-4}-10^{-2}$ becomes flatter than that of
the power-law distribution derived with the subsample \citep*[see
Fig. 5 in][]{2009MNRAS.397..135K} (note $\log L{\rm [OIII]}/M_{\rm
bh}\sim 1.7$ corresponding to the Eddington rate). In this work, we
use a power-law Eddington ratio distribution with an exponential
cutoff at a high ratio in all our calculations in order to avoid
inducing additional parameters, which can simulate the observed
Eddington ratio distributions quite well \citep*[see][for the
details]{2009ApJ...698.1550H}. This distribution can also describe
the main feature of the observed Eddington ratio distribution for
the whole sample given in \citet{2009MNRAS.397..135K}. The
investigation on the evolution of massive black holes by adopting
more realistic Eddington ratio distribution (e.g., a
power-law$+$log-normal Eddington ratio distribution) will be carried
out in our future work. }

\vskip 1cm

\figurenum{6}
\centerline{\includegraphics[angle=0,width=8.5cm]{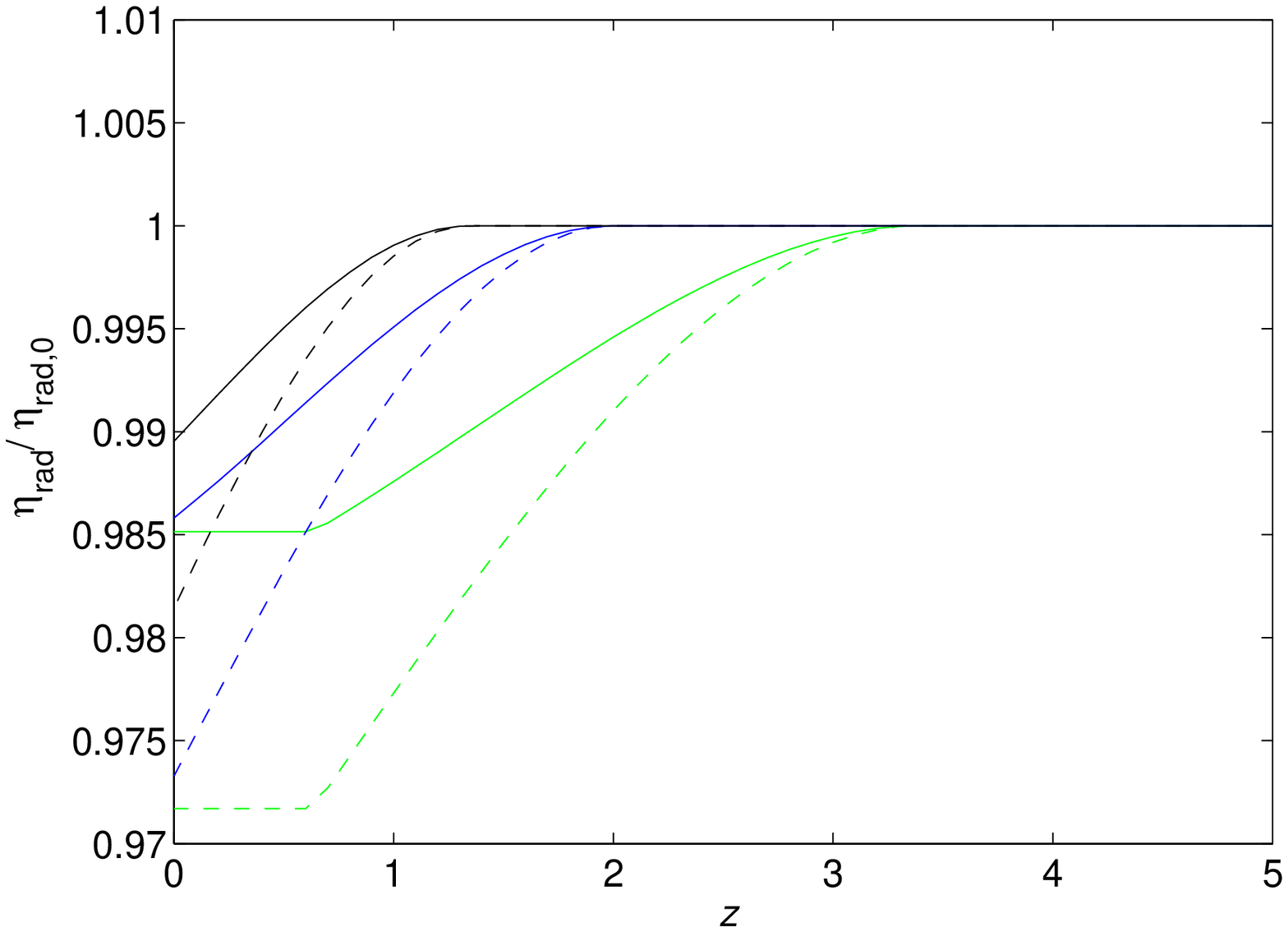}}
\figcaption{The radiative efficiencies as functions of redshift $z$.
The lines are derived with different quasar life timescales,
$\tau_{\rm Q}=2.5\times 10^8$ (green), $5\times10^8$ (blue), and
$7.5\times10^8$ years (black), respectively. The solid lines are the
results calculated with $s=0.5$, while the dashed lines are for
$s=1$. \label{fig_eta} }\centerline{}

In this work, we adopt a $\dot{m}$-dependent radiative efficiency
$\eta_{\rm rad}$, in which the accretion mode transition with mass
accretion rate $\dot{m}$ is considered. We perform calculations with
different values of $s$ (from $s=0$, i.e., a $\dot{m}$-independent
$\eta_{\rm rad}$, to $s=1$), and find that the final results are
insensitive to the value of $s$. {It is found that the radiative
efficiency evolves little with redshift (see Fig. \ref{fig_eta}).}
This is because the fraction of mass accreted in ADAF phases is
small compared with that in bright AGN phase
\citep{2005ApJ...631L.101C,2006ApJ...643..641H,2007ApJ...659..950C,2010ApJ...716.1423X}.
\citet{2000PASJ...52..133W}¡¯s calculations on the slim discs showed
that the radiative efficiency will not deviate significantly from
that for standard thin discs if $L_{\rm bol}/L_{\rm Edd}\la 2$,
which implies that the present adopted radiative efficiency
independent of Eddington ratio (\ref{etarad}) is indeed a good
assumption for the sources with $\dot{m}\ge\dot{m}_{\rm crit}$.

The Eddington ratio distribution is derived from a power-law quasar
lightcurve in this work. The situation becomes complicated at high
redshifts when the switch-on of AGN activity is not balanced with
switch-off of AGN activity, and the Eddington ratio distribution
(\ref{eddrat_dis}) may not be valid for the AGNs at high redshifts.
However, the black holes were dominantly grown up at $z\la 3$ (see
Fig. \ref{fig_bhmf_b}) when a power-law Eddington ratio distribution
is believed to be a good approximation. It has therefore not
affected much on our final results.

The comparison of the calculated BHMF of the AGN relics at $z=0$
with the local BHMF gives a lower limit on the quasar life timescale
$\tau_{\rm Q}$. The BHMFs of the AGN relics at $z=0$ are insensitive
to the adopted value of $\tau_{\rm Q}$, while they become
significantly different at relatively high redshifts (e.g., $z=1$,
see Fig. \ref{fig_bhmf_b}).  They can be compared with the measured
BHMF at $z$, which will set a further constraint on the quasar life
timescale $\tau_{\rm Q}$, provided the BHMFs are available at
relatively high redshift $z$. This is beyond the scope of this work.






\acknowledgments I thank the referee for his/her helpful comments,
and Francesco Shankar for providing the data of the local black hole
mass function. This work is supported by the NSFC (grants 10773020,
10821302, and 10833002), the National Basic Research Program of
China (grant 2009CB824800), the Science and Technology Commission of
Shanghai Municipality (10XD1405000), the CAS (grant
KJCX2-YW-T03), and the CAS/SAFEA International Partnership Program
for Creative Research Teams.

{}

\end{document}